\documentclass[twocolumn,showpacs,preprintnumbers,amsmath,amssymb]{revtex4}
\usepackage{graphicx}
\usepackage{dcolumn}
\usepackage{bm}
\usepackage{amsmath}
\newcommand{\bef}{\begin{figure}}
\newcommand{\eef}{\end{figure}}

\newcommand{\be}{\begin{equation}}
\newcommand{\ee}{\end{equation}}
\newcommand{\bea}{\begin{eqnarray}}
\newcommand{\eea}{\end{eqnarray}}
\begin{document}

\title{Scaling of elliptic flow in heavy-ion collisions with the number of constituent quarks in a transport model}

\author{Subhash Singha$^{1}$ and Md. Nasim$^{2}$}
\affiliation{$^{1}$Kent State University, Ohio, USA;\\
$^{2}$University of California, Los Angeles, USA}

\begin{abstract}
We studied the number of constituent quark scaling (NCQ) behaviour of
elliptic flow ($v_{2}$) under the framework of A Multi-Phase Transport model (AMPT) at both top-RHIC
and LHC energies. The NCQ-scaling in $v_{2}$ holds at top-RHIC
energy with AMPT string melting version, while it breaks in Pb+Pb collisions at LHC energy using
the same framework. The breaking of NCQ-scaling at LHC energy has been studied by varying
the magnitude of parton-parton scattering cross-section and lifetime of hadronic cascade
as implemented in AMPT. We find that the breaking of NCQ scaling   in
Pb+Pb collisions at $\sqrt{s_{NN}}$ =2.76 TeV is independent of the magnitude of
parton-parton cross-section and the later stage hadronic
interactions. Further we observed that scaling holds in a small
collision system like Si+Si at $\sqrt{s_{NN}}$ = 2.76 TeV. 
We discussed that the breaking of NCQ scaling is possibly due to high
phase-space density of constituents quarks in Pb+Pb collisions at
$\sqrt{s_{NN}}$ = 2.76 TeV. 
\end{abstract}
\maketitle

\section{INTRODUCTION}
Relativistic heavy-ion collision experiments aim to
study the formation and evolution of a strongly interacting matter
called Quark Gluon Plasma (QGP) ~\cite{whitepapers}. Experiments at Brookhaven
Relativistic Heavy Ion Collider (RHIC) and at CERN Large Hadron
Collider (LHC) established the existence of such strongly interacting matter, which is expected to be
formed micro-seconds after the big-bang.\\
The elliptic flow parameter, $v_{2}$, which is defined as a
second harmonic coefficient of the azimuthal Fourier decomposition of
the momentum distribution of produced particles has been widely used as
an excellent tool for understanding the dynamics of the system formed
in the early stages of high-energy heavy-ion collisions
~\cite{hydro,hydro1,hydro2,hydro3,hydro4, hydro5, early_v2}. This flow
parameter $v_{2}$ is extracted by studying the correlation of produced
particles with respect to the reaction plane ($\Psi$) as, 
\begin{equation}
v_{2}=\langle\cos(2(\phi-\Psi))\rangle,
\end{equation}
where $\phi$ is the azimuthal angle of the produced particles~\cite{flow_method}. \\
Results from RHIC-experiments show that at low transverse momentum ($p_{T}$ $<$ 2 GeV/c),
there is a clear mass-ordering of $v_{2}$ among the identified
hadrons~\cite{flow_star, flow_phenix}. It is observed that at 
fixed $p_{T}$, heavier hadrons have smaller values of $v_{2}$ than the
lighter ones. Hydrodynamic calculations suggest that the interplay between radial and elliptic flow plays an
important role in determining the  mass-ordering of $v_{2}$ at low
$p_{T}$~\cite{hydro,hydro1,hydro2,hydro3,hydro4, hydro5}. Subsequent later stage
hadronic re-scattering can also distort $v_{2}$ at low $p_{T}$~\cite{hirano}. 
It is observed that in the intermediate-$p_{T}$ region (2.0 $< p_{T} <$ 4.0 GeV/c), the
$p_{T}$-differential $v_{2}$ of baryons and mesons form separate
groups~\cite{flow_star, flow_phenix}. Such a baryon-meson splitting in $v_{2}$ is 
successfully reproduced by  models where quark-coalescence
mechanism is considered to be the dominant process for hadronization in
this $p_{T}$-regime~\cite{ncq,ncq1}. When both $v_{2}$ and $p_{T}$ of identified hadrons are
divided by number of constituent quarks ($n_{q}$), all the hadrons follow an approximate scaling
behaviour. This is known as number of constituent
quark (NCQ) scaling. The origin of such scaling is interpreted as an
evidence for dominance of quark degrees of freedom in the early stages
of heavy-ion collision. Another way of representing NCQ scaling is to
plot $n_{q}$ scaled $v_{2}$ as a function of $(m_{T}-m_{0})/n_{q}$,
where $m_{T}$ is transverse mass and $m_{0}$ is the rest mass of
hadron.\\
Recent $v_{2}$ results from LHC~\cite{flow_alice} show similar trend
of mass-ordering among the identified hadrons at low $p_{T}$  ($<$ 3 GeV/c) and 
about 30$\%$ increase in radial flow than the top-RHIC energy. But in the intermediate $p_{T}$ region (3.0 $< p_{T} <$ 6.0 GeV/c), the $v_{2}$ results
do not seem to follow NCQ-scaling as observed in lower energy RHIC
experiments. The $v_{2}$ of identified hadrons at LHC energy deviates
from NCQ-scaling at a level of 20$\%$. This
observation has triggered theoretical debate over the
NCQ-scaling. \\
A Multi-Phase Transport (AMPT) model with string melting version
(which includes parton coalescence) has been used to
reproduce the observed NCQ-scaling in $v_{2}$ at top-RHIC energies~\cite{ampt}. In
this paper, we investigated the behaviour of NCQ-scaling both at top-RHIC and
LHC energies using the  framework of AMPT model to understand the
reason behind it's breaking at higher energies.\\
This paper is organized as follows. In section \textrm{II}, we briefly
discuss the AMPT model. In section \textrm{III}, we describe the
NCQ-scaling behaviour of $v_{2}$ of identified hadrons at top-RHIC and LHC
energies using the AMPT model (version 1.11). The results are summarized
in section \textrm{IV}.
\begin{figure}
\begin{center}
\includegraphics[scale=0.4]{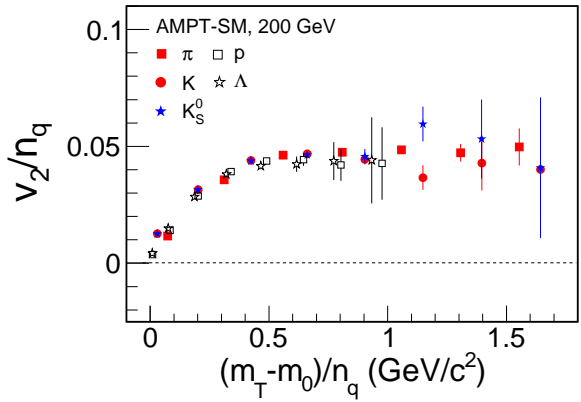}
\caption{(Color online) $v_{2}/n_{q}$ as a function of 
$(m_{T}-m_{0})/n_{q}$  for some selected hadrons ($\pi$, $K$,
$K^{0}_{S}$, $p$ and $\Lambda$) in minimum bias Au+Au collisions at
$\sqrt{s_{NN}}$ = 200 GeV using AMPT-SM model. The parton-parton cross-section is taken as
3 mb with hadronic cascade time = 30 fm in AMPT-SM model.}
\label{ncq_200}
\end{center}
\end{figure} 
\section{The AMPT Model}
The AMPT model, which is a hybrid transport model, has four main stages: the initial conditions,
partonic interactions, the conversion from the partonic to
the hadronic matter, and hadronic interactions~\cite{ampt}. It uses the same initial conditions as HIJING~\cite{hijing}.
Scattering among partons are modelled by Zhang's parton cascade~\cite{ZPC}, which calculates two-body parton scatterings
using cross sections from pQCD with screening masses. In the default
AMPT model, partons are recombined
with their parent strings and when they stop interacting,
the resulting strings fragment into hadrons according to the Lund
string fragmentation model~\cite{lund}. However,
in the string melting scenario (labeled as AMPT-SM), these strings are converted
to soft partons and a quark coalescence
model is used to combine partons into hadrons. The
evolution dynamics of the hadronic matter is described
by A Relativistic Transport (ART) model.  The interactions between the
minijet partons in the AMPT Default model and those between partons in the
AMPT-SM could give rise to substantial $v_2$. 
The parton-parton interaction cross section
in the string-melting version of the AMPT is taken to be  3 mb and 10
mb. In this study,
approximately 500 K (50 K) events for each configuration were
generated for minimum-bias Au+Au (Pb+Pb) collisions.
\begin{figure*}
\begin{center}
\includegraphics[scale=0.7]{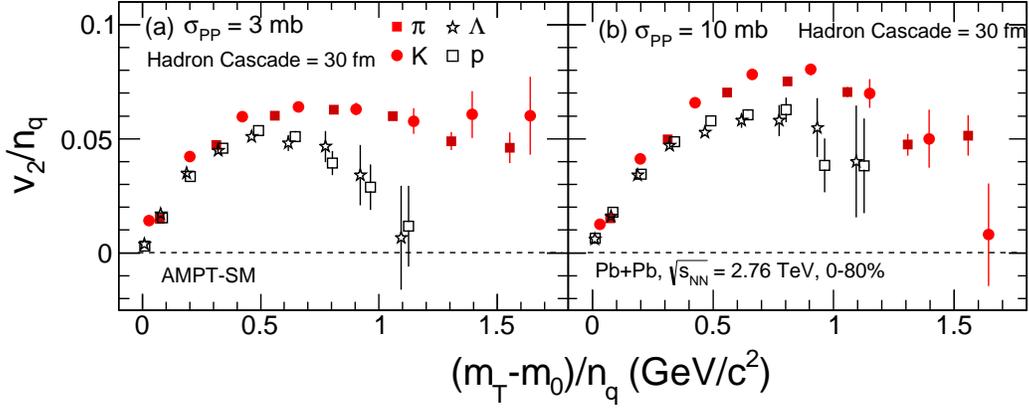}
\caption{(Color online) $v_{2}/n_{q}$ as a function of 
$(m_{T}-m_{0})/n_{q}$  for some selected hadrons ($\pi$, $K$,
 $p$ and $\Lambda$) in minimum bias (a) Au+Au collisions at
$\sqrt{s_{NN}}$ = 200 GeV  (b) Pb+Pb  collisions at  2.76 TeV  using AMPT-SM model. }
\label{ncq_276_150}
\end{center}
\end{figure*} 
\section{Results and Discussion}
It has been observed that NCQ scaling in $v_{2}$ holds for AMPT with
string melting scenario, which incorporates partonic coalescence
mechanism, but no such scaling occurs in the default AMPT~\cite{BN}. We studied the energy dependence of such
scaling using  AMPT-SM, mainly at top RHIC and LHC
energies. Fig~\ref{ncq_200} shows $v_{2}/n_{q}$ as a function of 
$(m_{T}-m_{0})/n_{q}$  for some selected hadrons ($\pi$, $K$,
$K^{0}_{S}$, $p$ and $\Lambda$) in minimum bias Au+Au collisions at
$\sqrt{s_{NN}}$ = 200 GeV using AMPT-SM model. A clear scaling is observed among all
hadrons consistent with the observation in Ref~~\cite{BN}. Here we used
parton-parton cross-section ($\sigma_{PP}$) equal to 3 mb and hadron
cascade time ($\tau$) equal
to 30 fm in these results.\\ 
After observing a clear scaling at 200 GeV,
we studied NCQ scaling in Pb+Pb collisions at  $\sqrt{s_{NN}}$ =
2.76 TeV using AMPT-SM model as shown in Fig.~\ref{ncq_276_150}. A clear breaking of scaling is observed for
$(m_{T}-m_{0})/n_{q}$ $>$ 0.4 GeV/$c^{2}$, which is
very striking and interesting as we have used AMPT-SM
model. Fig.~\ref{ncq_276_150}(a) and ~\ref{ncq_276_150}(b) show
scaling results where the magnitude of 
$\sigma_{PP}$ has been taken as 3 mb and 10 mb, respectively, keeping
same hadron cascade time (30 fm). It is clear that scaling
breaks down for both the values of $\sigma_{PP}$. This indicates that the breakdown
of NCQ scaling at  $\sqrt{s_{NN}}$ = 2.76 TeV is independent of
magnitude of parton-parton cross-section.\\
\begin{figure}[!h]
\begin{center}
\includegraphics[scale=0.4]{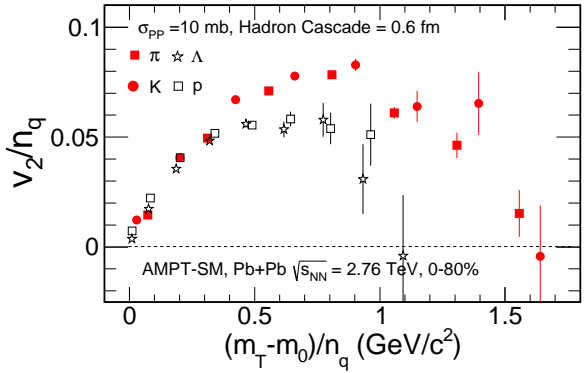}
\caption{(Color online) $v_{2}/n_{q}$ as a function of 
$(m_{T}-m_{0})/n_{q}$  for some selected hadrons ($\pi$, $K$,
 $p$ and $\Lambda$) in minimum bias Pb+Pb  collisions at  2.76 TeV
 using AMPT-SM model ($\sigma_{PP}$= 10 mb, $\tau$ = 0.6 fm).  }
\label{ncq_276_3}
\end{center}
\end{figure} 
One possible reason for the violation may be the distortion of  initially developed
$v_{2}$ by later hadronic
interaction. To check this effect, we turn-off hadronic cascade
in AMPT model. This can be done by setting input parameter
Nt=3, which gives hadron cascade time equal to 0.6 fm (minimum hadron cascade
time in AMPT). The NCQ scaling result from AMPT-SM ($\sigma_{PP}$= 10 mb) with hadron cascade
time 0.6 fm is shown in Fig.~\ref{ncq_276_3}.
In this case too we have observed that the scaling breaks, indicating that it is not due to the hadronic interactions at $\sqrt{s_{NN}}$= 2.76 TeV.\\
\begin{figure*}[!ht]
\begin{center}
\includegraphics[scale=0.85]{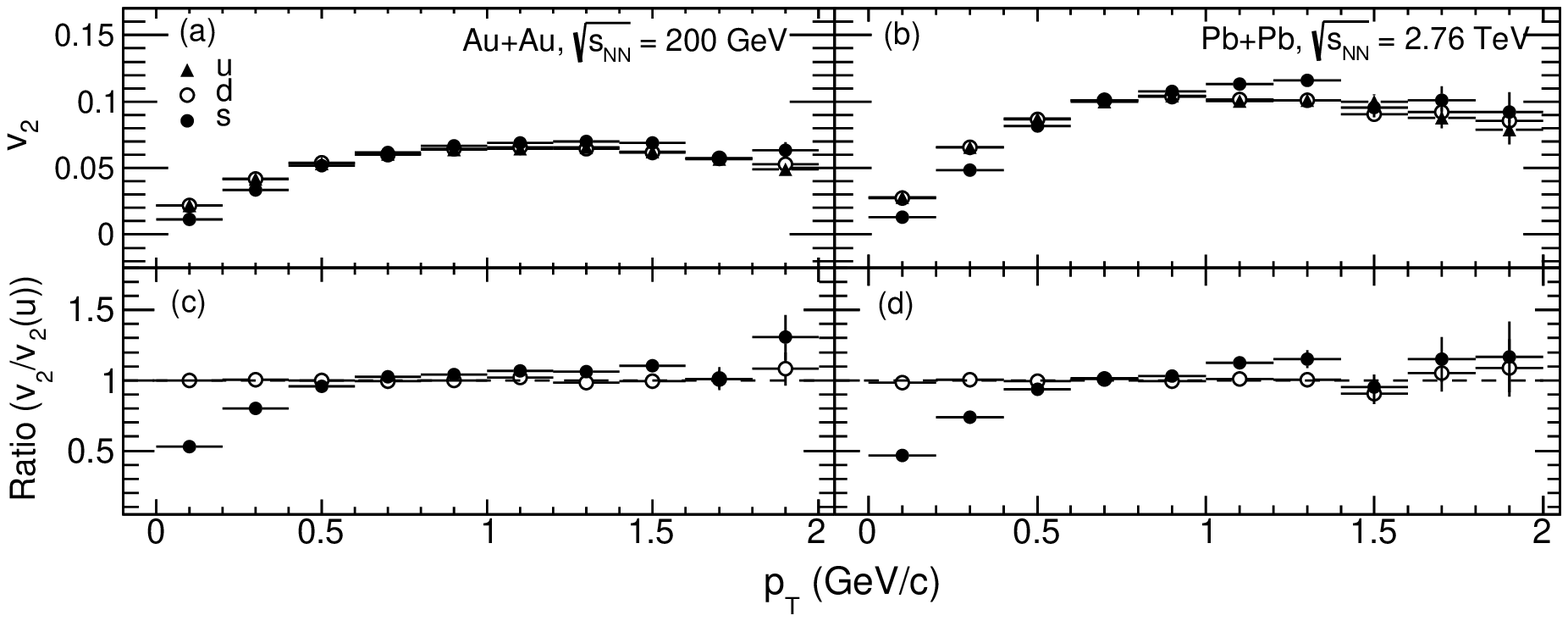}
\caption{(Color online) The $v_{2}$ of $u$, $d$ and $s$  quarks as a function of $p_{T}$ 
in AMPT-SM model ($\sigma_{PP}$ = 3 mb) for  minimum
bias  (a)  Au+Au collisions at $\sqrt{s_{NN}}$ = 200 GeV and  (b)
Pb+Pb  collisions at  2.76 TeV.  Ratios with respect to $u$-quark $v_{2}$ are shown in corresponding lower panel.}
\label{quarkv2}
\end{center}
\end{figure*} 
\subsection{Quark-$v_{2}(p_{T})$ distributions  in AMPT model}
According to coalescence model the
relation between  quark-$v_{2}$ ($v_{2}^{q}$) and hadrons-$v_{2}$
($v_{2}^{h}$) is as follows:
\begin{equation}
v_{2}^{h}(p_{T})=n_{q}v_{2}^{q}(p_{T}/n_{q}).
\label{coal}
\end{equation}
Where $p_{T}$ is the transverse momentum of hadron.
The violation of NCQ scaling at LHC energy within a parton coalescence
approach was first predicted in Ref~\cite{daniel}. According to  Ref~\cite{daniel}, modifications of the
underlying light and heavy quark  $v_{2}(p_{T})$ due to the strong
transverse expansion at LHC energy  could be the reason for NCQ
scaling violation.  
To understand such behavior in Pb+Pb collisions at 2.76 TeV ,
we have checked underlying $v_{2} (p_{T})$  for different quark
flavors in AMPT-SM model. \\
The $v_{2}$ of $u$, $d$ and $s$  quarks as a function of $p_{T}$
in the AMPT-SM model ($\sigma_{PP}$ = 3 mb)
are shown in Fig.~\ref{quarkv2} (a) and ~\ref{quarkv2} (b) for
$\sqrt{s_{NN}}$ = 200 GeV (Au+Au) and 2.76 TeV (Pb+Pb), respectively.
Ratios with respect to $u$-quark $v_{2}$ are shown in the corresponding
lower panel. 
 We have observed that the $v_{2} (p_{T})$ of $u$ , $d$ and $s$
 quarks are the same for both the energies in AMPT-SM model. However,
 for $p_{T}$ $<$ 0.5 GeV/c, magnitude of $s$-quark $v_{2}$ is
 slightly lower than that of $u$ and $d$. It is clear from
 Fig.~\ref{quarkv2}  that the $v_{2}(p_{T})$ distribution for
 different quark flavors is similar for both Au+Au and Pb+Pb collisions
 at  $\sqrt{s_{NN}}$ = 200 GeV and 2.76 TeV, respectively.  Therefore,
 the breaking of NCQ scaling  in AMPT-SM model for Pb+Pb collision at
 $\sqrt{s_{NN}}$ = 2.76 TeV is not due to change in $v_{2}(p_{T})$ of underlying quarks.

\subsection{Effect of parton density in coalescence mechanism}
\begin{figure}
\begin{center}
\includegraphics[scale=0.4]{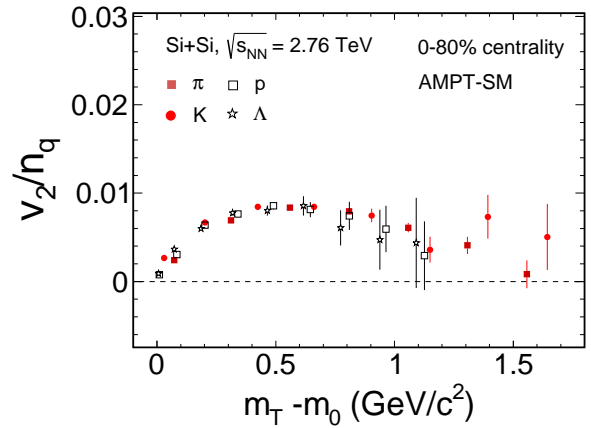}
\caption{(Color online) $v_{2}/n_{q}$ as a function of 
$(m_{T}-m_{0})/n_{q}$  for some selected hadrons ($\pi$, $K$,
 $p$ and $\Lambda$) in minimum bias Si+Si  collisions at
 $\sqrt{s_{NN}}$ = 2.76 TeV  using AMPT-SM model ($\sigma_{PP}$ = 3 mb,
 $\tau$ = 0.6 fm). }
\label{ncq_sisi}
\end{center}
\end{figure}

Let us recall the formalism of coalescence mechanism. In a simplified coalescence scenario, the probability that the constituents $a$ and $b$ will
form a composite object $C$~\cite{ncq} is 
\begin{equation}
  \begin{split}
    f_{C}(P_{C}, R, t_{c}) \approx  
    f_{a}(m_{a}P_{C}/(m_{a}+m_{b}), R, t_{c}) \\
\times  f_{b}(m_{b}P_{C}/(m_{a}+m_{b}), R, t_{c}).
    \label{coal_prob}
  \end{split}
\end{equation}
Here $f_{i}$ denotes phase densities, $P_{C}$ is the momentum of
the composite particle, $t_{c}$ is the coalescence time and $R$ is the centre-of-mass.
Masses of constituents are denoted by $m_{i}$.  Within the regime of coalescence mechanism, 
the invariant spectrum of produced particles is proportional to the product of the invariant spectra of constituents. 
Therefore, the yields of mesons and baryons produced by coalescence of quarks (q)  are given by
\begin{equation}
\frac{dN_{B}}{d^{2}p_{T}}(p_{T}) = f_{B}(p_{T})[ \frac{dN_{q}}{d^{2}p_{T}} (p_{T}/3) ]^{3} 
\label{coal_yieldB}
\end{equation}
\begin{equation}
\frac{dN_{M}}{d^{2}p_{T}}(p_{T}) = f_{M}(p_{T})[ \frac{dN_{q}}{d^{2}p_{T}}(p_{T}/2)  ]^{2},
\label{coal_yieldM}
\end{equation}
where the coefficient $f_{M}$ and $f_{B}$ are the probabilities  for meson and baryon coalescence.
Note that Eq.~\ref{coal},~\ref{coal_yieldB} and~\ref{coal_yieldM},  are valid only when the phase space density is very small~\cite{ncq}. 
When phase-space density of quarks is very high,  the probability to
find another quark in vicinity will be close to unity. So the final
composite $v_{2}$ of hadron will be linear in terms of the quark's
$v_{2}$ and hence breaking the scaling relation. 
On the other hand for low density, a quark has a small probability of
finding another quarks to coalesce, and Eq.~\ref{coal},~\ref{coal_yieldB} and~\ref{coal_yieldM} will be valid.\\
So the change in phase space density of quarks can affect coalescence
mechanism  and it can be studied using AMPT model.
We  generated 2 million Si+Si collision events at $\sqrt{s_{NN}}$ = 2.76 TeV using the same AMPT-SM
configuration  ($\sigma_{PP}$= 3 mb, $\tau$ =0.6 fm). Because of small system size, we would expect a
smaller density compared to that in  Pb+Pb collisions. So if NCQ-scaling
at LHC energies in Pb+Pb collisions breaks due to high density of partons, the
scaling might hold in Si+Si collision system at same centre-of-mass energy.
Fig.~\ref{ncq_sisi} shows the NCQ scaling plot for minimum bias Si+Si
system at  $\sqrt{s_{NN}}$ = 2.76 TeV.  We can see that NCQ scaling holds much better than Pb+Pb system.
This confirms that the breaking of NCQ-scaling of $v_{2}$  in Pb+Pb
collisions at  $\sqrt{s_{NN}}$ = 2.76 TeV is due to very high
phase-space  density  of initially produced quarks.

\section{Summary and Conclusion}
In summary, we have studied the number of constituent quark scaling in
$v_{2}$ for hadrons at top-RHIC and LHC energies using AMPT-SM
model.  We have observed that while NCQ-scaling holds at $\sqrt{s_{NN}}$
= 200 GeV but model fails to reproduce the same in Pb+Pb collisions at
$\sqrt{s_{NN}}$ = 2.76
TeV. We have observed the breaking in NCQ scaling at $\sqrt{s_{NN}}$
= 2.76 TeV is independent of the magnitude of parton-parton
cross-section and also not due to later stage hadronic interactions.
We also compared $v_{2}$ of $u$, $d$ and $s$  quarks as a function of
$p_{T}$ for Au+Au and Pb+Pb collisions in AMPT-SM model to see any possible
change in underlying  quark  $v_{2}(p_{T})$ due large radial flow at
LHC energy. We find $v_{2}(p_{T})$ of $u$, $d$ and $s$  quarks shows
similar behaviour for both Au+Au and Pb+Pb collisions. Therefore, the
violation in NCQ scaling is not due to change in  underlying  quark
$v_{2}(p_{T})$  in Pb+Pb collisions at LHC energy. 
Further we checked the effect of parton's phase-space density on NCQ
scaling behaviour within the framework of coalescence.
We observed that the scaling holds in a small collision system like Si+Si
at $\sqrt{s_{NN}}$ = 2.76 TeV where the phase-space density of
constituent quarks is not very high as compared to Pb+Pb . This observation can be well understood
in the framework of coalescence mechanism. 
Our study shows that the NCQ-scaling in $v_{2}$ is not a necessary condition for quark coalescence when phase-space density of constituent quarks is very high, e.g Pb+Pb collision LHC energies.

\noindent{\bf Acknowledgments}\\
Financial support from DOE project, USA is gratefully acknowledged.
SS acknowledges support from the DOE project DE-FG02-89ER40531.

\end{document}